\documentclass[a4paper,11pt]{article}

\usepackage[a4paper,
  left=2.5cm, right=2.5cm,
  top= 3cm, bottom=4cm]{geometry}
\usepackage{amsmath}
\usepackage{oldgerm}
\usepackage{amssymb}
\usepackage{bbm}
\usepackage[dvips]{graphicx}
\usepackage{epsfig}
\usepackage{color}
\usepackage{cite}
\usepackage{epic}
\usepackage{hyperref}

\newcommand{\be}{\begin{equation}}  
\newcommand{\ee}{\end{equation}}

\newcommand{\rem}[1]{} 

\def\C{\mathbb{C}}
\def\Z{\mathbb{Z}}

\def\P{\mathbb{P}}

\def\I{\mathbb{I}}

\def\Hirz[#1]{\mathbbm{F}_{#1}}
\def\o[#1]{\overline{#1}}

\def\cF{\mathcal{F}}

\def\cC{\mathcal{C}}
\def\cD{\mathcal{D}}

\begin{document}

\begin{titlepage}

\vskip -0.5cm
\rightline{\small{\tt KCL-MTH-14-03}}

\begin{flushright}

\end{flushright}
 
\vskip 1cm
\begin{center}
 
{\large \bf Hypercharge flux in F-theory and the stable Sen limit} 
 
 \vskip 1.2cm
 
 A.~P.~Braun$^1$, A.~Collinucci$^2$ and R.~Valandro$^{3}$

 \vskip 0.4cm
 
 {\it $^1$Department of Mathematics, King's College, London WC2R 2LS, UK
\\[2mm]

 $^2$Physique Th\'eorique et Math\'ematique, Universit\'e Libre de Bruxelles, C.P. 231, \\1050
Bruxelles, Belgium \\[2mm]
 
 $^3$ICTP, Strada Costiera 11, Trieste 34014, Italy.
 }
 \vskip 1.5cm
 
\abstract{IIB compactifications enjoy the possibility to break GUT groups via fluxes without giving mass to the hypercharge gauge field. Although this important advantage has greatly motivated F-theory constructions, no such fluxes have been constructed directly in terms of the M-theory $G_4$-form.
In this note, we give a general prescription for constructing \emph{hypercharge G-fluxes}. By using a stable version of Sen's weak coupling limit, we verify their connection with IIB fluxes. We illustrate the lift of fluxes in a number of examples, including a compact ${\rm SU}(5) \times {\rm U}(1)$ model with explicit realization of doublet-triplet splitting. Finally, we prove an equivalence conjectured in an earlier work as a by-product.} 

\end{center}

\end{titlepage}

\tableofcontents

\newpage

\section{Introduction}

In model building applications, one of the most notable advantages of IIB compactifications over Heterotic ones is their ability to break GUT groups via fluxes without paying the price of giving mass to the hypercharge. This phenomenon was discovered in \cite{Buican:2006sn} in IIB string theory (see also \cite{Tatar:2008zj,Blumenhagen:2008zz}), and was propagated to F-theory in \cite{Beasley:2008dc,Beasley:2008kw,Donagi:2008ca,Donagi:2008kj,Marsano:2009ym,Dudas:2010zb,Marsano:2010sq,Palti:2012dd,Mayrhofer:2013ara} (see \cite{Weigand:2010wm,Maharana:2012tu} for a review).

The basic idea is very geometric: Take a CY threefold $X_3$, and place a stack of D7-branes on a divisor $D \subset X_3$, carrying some GUT gauge group. One could Higgs the GUT group. However, this requires control over the Higgs potential, which is lacking for compact models. An alternative is to turn on a DBI flux $F_2$, a two-form of $H^2(D)$, with values along a Cartan subgroup. One could construct $F_2$ as a two-form in $H^2(X_3)$, and restrict it to $D$. However, such a flux will generically give mass to the $U(1)_Y$ via the St\"uckelberg mechanism.\footnote{When $h^{1,1}(X_3)\neq 0$, a non-trivial (odd) flux can still leave the $U(1)_Y$ massless (see \cite{Mayrhofer:2013ara}).}

Fortunately, the cohomology of a divisor is typically much richer than that of the threefold in which it lives, $H^2(X_3, \Z) \subsetneq H^2(D, \Z)$. A sufficient condition on $F_2$ to keep $U(1)_Y$ massless is 
\begin{equation}
\int_D F_2 \wedge \imath^*\omega_2 = 0 \quad \forall \,\, \omega_2 \in H^2(X_3, \Z)\,,
\end{equation}
where $\imath$ is the inclusion map $\imath: D \hookrightarrow X_3$ (we will discuss the necessary and sufficient condition in section \ref{sec:HyperchFl}). This condition forbids $F_2$ from being the pullback of a two-form of $X_3$. However, this is not sufficient. The orthogonality condition implies that $F_2$ must be the Poincar\'e dual of a \emph{difference} of at least two holomorphic curves in $D$, $F_2 = \cC_1-\cC_2$. Moreover, each one of these curves must be Poincar\'e dual to a non-pullback two-form. 

To put it in general terms, $H^{1,1}(D)$ can be decomposed into a component that is pulled back from $X_3$, and its orthogonal complement: $H^{1,1}(D) = \imath^*H^{1,1}(X_3) \oplus(\imath^*H^{1,1}(X_3))_\perp$. However, \emph{integral} fluxes cannot be decomposed in this way. This decomposition is too coarse, and misses a crucial set $\cD$ of generators known as \emph{glue vectors.}\footnote{See \cite{Denef:2007vg} for an explanation in a similar context. In the present case, what is missed are elements in the quotient of $H^{1,1}(D)\cap H^2(D,\Z)$ by $\imath^*H^{1,1}(X_3)\cap H^2(D,\Z) \oplus(\imath^*H^{1,1}(X_3))_\perp\cap  H^2(D,\Z)$. These are precisely the lattice elements gluing the mutually orthogonal lattices $\imath^*H^{1,1}(X_3)\cap H^2(D,\Z)$ and $(\imath^*H^{1,1}(X_3))_\perp\cap  H^2(D,\Z)$ to form the full $H^{1,1}(D)\cap H^2(D,\Z)$.} Roughly speaking, for a divisor the form $P \equiv A\,B - C\,D = 0$, a typical glue vector will be a curve $\cC$ of the form
\begin{equation} \label{gluevector}
\cC: \quad A=0 \quad \cap \quad C=0 \qquad \subset \qquad D:\quad P \equiv A\,B-C\,D=0\,.
\end{equation}
A typical glue vector $\cC \in \cD$ cannot be written as a pulled back two-form by definition, but it will still not be orthogonal to all of the pulled back forms. A hypercharge flux must be written as a difference
\begin{equation}
F_Y = \cC_1-\cC_2 \quad {\rm for} \quad \cC_{1,2} \in \cD\,.
\end{equation}
This framework for GUT breaking has been understood since \cite{Buican:2006sn}. However, no single explicit analogue has been created in F-theory by way of G-fluxes. The most common approaches so far have been: Forgetting the CY fourfold, and constructing two-forms on a brane in the base of the elliptic fibration. Or, using the Heterotic/F-theory duality as a \emph{definition} of flux for F-theory.

In \cite{Mayrhofer:2013ara}, the authors explained what structure such a G-flux must have, in order to mimic the IIB notion of a hypercharge flux. The main requirement is that its Poincar\'e dual must be a difference of at least two four-cycles in the CY fourfold, each one a $\P^1$-fibration of a curve of type $\cC \in \cD$ as described above. Such a four-form cannot be written as the difference of two four-forms that factorize into wedge products of two-forms. 

In this paper, we give a general procedure for constructing such G-fluxes. The key insight is the following: After resolving the non-Abelian singularities over a brane at $P=0$, the fourfold is described by the Tate model, and an equation $\sigma v =P$, where $\sigma$ is an auxiliary coordinate, and $v=0$ is the exceptional divisor. It is then straightforward to find four-cycles $\gamma$ of the form:
\begin{equation}
\gamma: \quad v=0 \quad \cap \quad A=0 \quad \cap \quad  C=0 \qquad \subset \qquad \sigma\,v= A\,B-C\,D\,,
\end{equation}
which mimics the construction in \eqref{gluevector}. 

Although we are also able to construct our fluxes directly in F-theory, we will use a new technique that allows us to stay in touch with perturbative physics and check that we are on the right track. As a by-product, we will also show how to lift a typical flux on a so-called \emph{Whitney} brane, which carries only a $\Z_2$ gauge group. Using a stable version of Sen's weak coupling limit, given by Clingher, Donagi, Wijnholt \cite{Clingher:2012rg}, we give a
general procedure for lifting IIB hypercharge fluxes to G-fluxes in F-theory. 

In its original version, Sen's limit is only performed at the level of the discriminant, so that data such as fluxes, which live in the fourfold geometry, cannot be directly discussed. Applying Sen's limit directly to the fourfold drastically mutilates the geometry, thereby losing most of the 7-brane data. In short, we do not have a weakly coupled F-theory fourfold \emph{per se}.

The stable version of Sen's limit introduced in \cite{Clingher:2012rg} satisfyingly addresses the issue. The basic idea is to consider the whole family of CY fourfolds over the parameter $\epsilon$ controlling the limit to weak coupling. This object is a Calabi-Yau fivefold with a singularity at $\epsilon=0$. Blowing up the singular locus, the fourfold over $\epsilon=0$ splits into two components. One component only sees perturbative physics, i.e. the only monodromies acting on the elliptic fiber are $T$ and $-\I$  $\in {\rm SL}(2, \Z)$. The other component adds the non-perturbative phenomena. This clean way of geometrizing the weak coupling limit of F-theory allows us to directly lift fluxes form IIB to F-theory.

This paper is organized as follows: In section \ref{sec:Senlimit}, we review the new formulation of Sen's limit. In section \ref{sec:LiftF2G4} the limit is used to lift IIB two-form fluxes to F-theory four-form fluxes. Section \ref{sec:HyperchFl} contains our main results: we show how to explicitely construct fluxes that leave the corresponding gauge field massless in F-theory and apply this to an $SU(5)\times U(1)$ model. In section \ref{sec:u1resSmW} we prove a conjectured equivalence between two-form and four-form fluxes given in \cite{Braun:2011zm,Grimm:2010ez}.


\section{The stable version of Sen's weak coupling limit} \label{sec:Senlimit}


Supersymmetric F-theory compactifications to four dimensions require a Calabi-Yau fourfold that is elliptically fibered over a base manifold $B_3$ as part of the defining data. 
If the elliptic fibration has a section, the fourfold can be described by a Weierstrass model:
\begin{equation} \label{WM}
y^2=x^3 + x z^4 f+z^6 g \:.
\end{equation}
Here, $x,y,z$ are taken to be sections of $(\bar{K}_B\otimes F)^{\otimes 2}$,  $(\bar{K}_B\otimes F)^{\otimes 3}$, respectively. The fiber coordinates are embedded into
$\P^2_{123}$ and $F$ is the line bundle associated with the hyperplane section of that space. $K_B$ is the canonical bundle of the base $B_3$. 
It follows that $f$ and $g$ are sections of $\bar{K}_B^{\otimes 4}$ and $\bar{K}_B^{\otimes 6}$, respectively. We can express them in terms of the 
sections $a_i\in \bar{K}^{\otimes i}$ appearing in the Tate form: 
\begin{equation}\label{fgWCL}
 \begin{array}{l}
 f=-\frac{b_2^2}{3}+2b_4 \\ \\
g=\frac{2}{27}b_2^3-\frac23 b_2b_4+b_6 \\
 \end{array} \qquad \qquad \mbox{where} \qquad \qquad
 \begin{array}{l}
 b_2=a_2+\frac14 a_1^2 \\
b_4=\frac12 a_4 + \frac14 a_1a_3 \\
b_6=a_6 + \frac14 a_3^2 \\
 \end{array} \:.
\end{equation}

A refined version of Sen's weak coupling limit appeared recently in \cite{Clingher:2012rg,Donagi:2012ts}. The starting point is the same as Sen's and consists in scaling the $b_i$'s with a parameter $\epsilon$ that controls the limit:
\begin{equation}
 b_2 \rightarrow \epsilon^0 \, b_2\, , \qquad b_4 \rightarrow \epsilon^1\, b_4\, , \qquad b_6 \rightarrow \epsilon^2 \, b_6 \:.
\end{equation}
When $\epsilon \rightarrow 0$, the $j$-function of the elliptic fiber goes to zero as $\epsilon^2$ away from the vicinity of $b_2=0$ and, correspondingly, the string coupling becomes small almost everywhere over the base space $B_3$. From the perspective of the geometry of the fourfold of F-theory, however, this limit leads to a severe singularity \cite{Clingher:2012rg,Donagi:2012ts}. If we introduce the coordinate $s=x-\frac{1}{3}b_2z^2$ and rewrite the Weierstrass equation by using the parametrization \eqref{fgWCL} for $f$ and $g$ we find
\begin{equation}\label{W5sing}
 y^2 = s^3 + b_2 s^2\, z^2+ 2b_4 s\,\epsilon\,z^4 + b_6 \epsilon^2\,z^6 \:.
\end{equation}
This is a family of Calabi-Yau fourfolds over the $\epsilon$-plane. At $\epsilon=0$, the elliptic fiber degenerates over all points of $B_3$. What is worse $b_4$ and $b_6$, i.e.
the information on the location of the D7-brane locus, is lost completely. 

To properly understand this degeneration it is necessary to consider the whole family of fourfolds described by \eqref{W5sing}. The resulting fivefold $W_5$ is singular at $y=s=\epsilon=0$, i.e. the degeneration is not of a stable type. By blowing up the singular locus of the whole family\footnote{A similar procedure is needed to describe 
the stable degeneration limit of a K3 surface which forms the basis for the duality between F-theory and heterotic string theory.} this situation can be improved and the perturbative information about the D7-brane locus (i.e. information about a neighbourhood of the locus $\epsilon=0$ in the family $W_5$) is recovered in the limit $\epsilon\rightarrow 0$. 
The resolved fivefold is given by
\begin{equation}
\tilde{W}_5:\quad y^2=s^3\lambda + b_2 s^2\, z^2+ 2b_4 s\,t\, z^4 + b_6 t^2\,z^6 \:,
\end{equation}
i.e. it is an hypersurface in the ambient sixfold 
\begin{center}
\begin{tabular}{ccccc}
$y$ & $s$ & $t$ & $z$ & $\lambda$ \\
$3$ & $2$ & $0$ & $1$ & $0$ \\
$1$ & $1$ & $1$ & $0$ & $-1$ \\
$3\bar{K}_B$ & $2\bar{K}_B$ & $0$ & $0$&$0$
\end{tabular}\\
\end{center}
with an SR ideal generated by $[syz],[yst],[z\lambda]$. The blow up $\tilde{W}_5\rightarrow W_5$ is given as the inverse of the 
map $s\mapsto s \lambda$, $y\mapsto y\lambda$, $\epsilon \mapsto t\lambda$. 

The type IIB data (i.e. the weak coupling limit) is captured by the central fiber at $\epsilon=0$ of the family $\tilde{W}_5$. 
As $\epsilon = t\lambda$, the fourfold splits up into two components: $X_4^0=W_T \cup_{X_3} W_E$. Here
\begin{align}
 W_T:\quad \tilde{W}_5\cap\{t=0\}:\quad & y^2=s^2(b_2z^2+s\lambda)\\
 W_E:\quad \tilde{W}_5\cap\{\lambda=0\}:\quad &   y^2=b_2 s^2\, z^2+ 2b_4 s\,t\, z^4 + b_6 t^2\,z^6 
\end{align}
The two fourfolds $W_E$ and $W_T$ intersect along the threefold 
\begin{equation}
 X_3:\quad \tilde{W}_5\cap \{t=0\}\cap\{\lambda=0\}:\quad y^2=z^2s^2 b_2 \, .
\end{equation}
Since $[yst]$ and $[\lambda z]$ belong to the SR ideal, we can define
$\xi=y/(sz)$ and write $X_3$ as 
\begin{equation}\label{X_3}
 X_3 \,: \qquad \xi^2=b_2 \, .
\end{equation}
One immediately recognizes this as a Calabi-Yau double cover $X_3$ of $B_3$. This procedure allows us to find the type IIB Calabi-Yau threefold 
as a submanifold of the (degenerate) F-theory fourfold over $\epsilon=0$.

In the following, we will only be interested in $W_E$. As it is defined by $\lambda=0$, we can exploit the SR-ideal to set $z=1$, which is assumed from now on. 
$W_E$ can then be written as
\begin{align}\label{WEeq}
 W_E:\quad &   y^2= b_2s^2+2b_4s\,t+b_6 t^2 
\end{align}
in an ambient fivefold $Y_5$ which is spanned by the base manifold $B_3$ and the three homogeneous coordinates $(s:y:t)$ (with equal weight $1$). Hence, \eqref{WEeq} 
is a quadratic equation in a $\P^2$, in other words it is a conic bundle, the generic fiber of which is a $\P^1$. This $\P^1$ fiber degenerates into two $\P^1$'s over the discriminant locus of the quadratic equation, i.e. this happens when
\begin{equation}
\Delta_E\equiv  \det\left(\begin{array}{cc}b_2 & b_4 \\ b_4 & b_6 \end{array}\right)=0 \quad\mbox{in} \, B_3 \, .
\end{equation}
We recognize this as the locus of the $D7$ brane in $B_3$. The two rational curves fibered over the subset $\Delta_E\subset B_3$ make up a threefold $R_3\equiv W_E\cap\{\Delta_E=0\}$, that we call the cylinder (even if the proper cylinder is a normalization of $R_3$ \cite{Clingher:2012rg}). This is the basic object needed to lift fluxes on D7-branes as we will see in the next section.

\section{Lifting type IIB $\cF_2$ fluxes to F-theory $G_4$ fluxes}\label{sec:LiftF2G4}

In this section we present the basic ideas in simple non-compact examples. 

\subsection{One brane and its image}

We consider Type IIB string theory on an orientifold of $\mathbb{R}^{1,3}\times X_3$, where the Calabi-Yau threefold $X_3=\C^3$ is described by the following equation in $\C^4$ (with coordinates $x_1,x_2,x_3,\xi$):
\begin{equation}\label{localCY3}
  X_3\,: \qquad \xi^2 = x_3 + 1 \:.
\end{equation}
The orientifold involution is $\xi\mapsto -\xi$, whose quotient is $B_3=\C^3$ with coordinates $x_1,x_2,x_3$. The orientifold plane is located at $b_2\equiv x_3+1=0$ in $B_3$. We now place one brane on the locus $\xi-1 =0$ and its image on $\xi+1=0$. This choice corresponds to $b_4=1$ and $b_6=1$ (we can make this simple choice because we are in a non-compact setup). We can now construct the conic bundle $W_E$ over $B_3$ corresponding to this brane setup:
\begin{equation}
  W_E\,: \qquad y^2 = (x_3+1)s^2 + 2 s\,t + t^2 \:.
\end{equation} 
Note that the fiber is compact: it is a $\P^1$ that degenerates over the brane locus $\Delta_E\equiv x_3=0$.
The conic bundle can be rewritten in the following form:
\begin{equation}
  W_E\,: \qquad (y-s-t)(y+s+t) = x_3s^2 \:.
\end{equation} 
We note that the fourfold $W_E$ is smooth (the apparent conifold singularity at $y=s=t=x_3=0$ is excluded by the SR-ideal). This is consistent with the fact that the two branes intersect each other only on top of the O7-plane where the matter is projected out. The Calabi-Yau threefold $X_3$ is embedded in $W_E$ by $t=0$ ($\xi\sim y/s$). 

The cylinder $R_3=W_E \cap \Delta_E$ splits into two components $R_3^\pm$: $x_3=y\pm(s+t)=0$. This means that, on top of the locus $\Delta_E\equiv x_3=0$, the conic fiber splits into two $\P^1$'s which we call $\P^\pm_{\Delta}$. They are described by the equations $y\pm(s+t)=0$ in the $\P^2$ with coordinates $y,s,t$. The fiber over $\Delta_E=0$ can be pictured as
\begin{equation}
   \P^+_\Delta \cup_{p} \P^-_\Delta
\end{equation}
where $p$ is the intersection point $p:\{y=t+s=0\}$ between the two $\P^1$'s. The vicinity of this point in $W_E$ looks like a Taub-NUT space:
\begin{equation}
 y_+\, y_- = x_3 \:,
\end{equation}
where we have used that $p$ is in the patch $s=1$ and we have defined $y_\pm \equiv y\pm(s+t)$. M-theory compactified on Taub-NUT is the lift of a type IIA background with a D6-brane. The D6-brane lies where the M-theory circle $S^1_M$ collapses, in this case, at $x_3=0$. The direction orthogonal to $S^1_M$ in the fiber forms another circle when this Taub-NUT is embedded into a compact elliptic fibration. By T-dualizing it, the D6-brane becomes a D7-brane along $x_3=0$ in $B_3$. The weak-coupling limit splits the elliptic fourfold into a $\P^1$ and a `compactified' Taub-NUT space. In this simple example, the double cover of the locus $\Delta_E=0$ splits into a brane/image-brane system.

Given a two-form flux $\cF$ on the D6-brane wrapping the four-cycle $\Delta_E=0$ in $B_3$, it is known how to lift it to a four-form flux $G_4$ on the fourfold $W_E$ \cite{Gubser:2002mz}: one takes the Poincar\'e dual two-cycle of $\cF$ on the brane, say $\cC_{\cF}$. Above this curve in $B_3$ the conic splits into two $\P^1$'s. We can fiber one of the two $\P^1$'s over the curve $\cC_{\cF}$, thereby lifting $\cC_{\cF}$ either in $R_3^+$ or in $R_3^-$. These two six-cycles sum up to a `pull-back' cycle (i.e. given by the intersection of a divisor of the ambient space with the fourfold $W_E$).
The corresponding four-form flux is, up to a four-form of the pull-back type ( $\sim$ ), given by
\begin{equation}
 G_4 \sim \cF\wedge R_3^+ \sim - \cF \wedge R_3^- \sim \cF \wedge \frac12 \left( R_3^+ - R_3^- \right) \:. 
\end{equation}
In F-theory, we want to impose 4d Poincar\'e invariance (i.e. the flux must be orthogonal to base and fiber \cite{Dasgupta:1999ss}), for which the right definition turns out to be the last one, i.e.
\begin{equation}\label{G4cFlift}
 G_4 = \frac12 \, \cF \wedge  \left( R_3^+ - R_3^- \right) \:.
\end{equation}
This flux projects to $\cF$ on the brane $\xi+1=0$ and to $-\cF$ on the brane $\xi-1$, realizing an orientifold invariant fluxed brane configuration.

\

This conic bundle comes from the following family of elliptic fibrations over $B_3$:
\begin{equation}
 W_5\,: \qquad (y-s\,z-\epsilon\,z^3)(y+s\,z+\epsilon\,z^3) = (s+x_3z^2)s^2 \:.
\end{equation}
We notice that there is a new section of the elliptic fibration (with respect to the most generic elliptic fibration over $B_3$), i.e. $\sigma_\pm$: $y\pm z^3\epsilon=s=0$. 

The $G_4$ flux \eqref{G4cFlift} is not defined in a generic elliptic fourfold of the family $W_5$. We need to find an object in its homology class in $W_E$ that is well defined also away from weak coupling. The sections $\sigma_\pm$ are manifestly well defined on each element of the family $W_5$. They satisfy the homological relation
\begin{equation}
  [y_\pm]|_{W_E} = R_3^+ + 2\sigma_+ =  R_3^- + 2\sigma_- 
\end{equation}
with the $R_3^\pm$. We can then define the flux as
\begin{equation}\label{G4cFlift2}
 G_4 = \cF \wedge  \left( \sigma_- - \sigma_+ \right) \:.
\end{equation}
This is in accordance with the expectation for the ${\rm U}(1)$ restriction (in this case $b_6$ is a square).

\subsection{Non-abelian gauge group: ${\rm SU}(2)$ stack} \label{Sec:NnAbSU2loc}

Let us consider a different brane setup on the Calabi-Yau threefold \eqref{localCY3}, supporting a ${\rm U}(2)$ gauge group. We place the stack of two branes (plus their images) at $x_3=0$. This is realized by choosing $b_4=x_3$ and $b_6=0$ and correspondingly $\Delta_E=x_3^2$. The related conic bundle is
\begin{equation}
W_E\,: \qquad (y-s)(y+s)=s \,x_3 (s + 2 t)
\end{equation}
This has an $A_1$ singularity along $y=s=x_3=0$, which we resolve in the following way. We introduce a new coordinate $\sigma$ and a new (auxiliary) equation $\sigma=x_3$ \cite{Collinucci:2010gz}. The resolved fourfold is then described as follows 
\begin{equation}
\tilde{W}_E\,: \qquad \left\{ \begin{array}{l} (y-s)(y+s)=s \,\sigma\, (v\,s + 2 t) \\  \sigma\,v=x_3 \\ \end{array} \right.
\end{equation}
In the resolved space, the fiber coordinates $y,s,t,\sigma,v$ satisfy the scaling relations summarized in the following table
\begin{center} 
\begin{tabular}{ccccc}
$y$ & $s$ & $t$ & $\sigma$ & $v$ \\
$1$ & $1$ & $1$ & $0$ & $0$ \\
$1$ & $1$ & $0$ & $1$&$-1$ \\
\end{tabular}\\
\end{center}
The SR-ideal is $[yst],[ys\sigma],[vt]$.

The cylinder is $\hat{R}_3=\hat{W}_E\cap \{x_3=0\}$. It splits into three components
\begin{equation}
R^+_{\sigma}\,: \,\,\, \left\{ \begin{array}{l}
y + s =0  \\  \sigma=0  \\ x_3=0 \\
\end{array}\right.  \qquad
R^-_{\sigma}\,: \,\,\, \left\{ \begin{array}{l}
y - s =0  \\  \sigma=0  \\ x_3=0 \\
\end{array}\right.  \qquad
R_{v}\,: \,\,\, \left\{ \begin{array}{l}
y^2=s^2+ 2\,s\, t \sigma \\  v=0 \\ x_3=0 \\
\end{array}\right.
\end{equation}
On top of a generic point of the locus $\Delta_E=0$, the conic splits into three $\mathbb{P}^1$'s, that we call $\P_\sigma^+$, $\P_\sigma^-$ and $\P_v$ (these are the fibers of $R_\sigma^+$, $R_\sigma^-$ and $R_v$, respectively). Their intersection structure is schematically 
\begin{equation}
  \P_\sigma^+ \,\, \cup_{p_1} \,\, \P_v \,\, \cup_{p_2} \,\, \P_\sigma^-
\end{equation}
In the M-theory picture, the resolved fiber can be understood as the result of separating the two D6-branes of the ${\rm U}(2)$ stack. 
We will take a two-form flux $\cF_1$ on the first brane (located at $p_1=\P_\sigma^+ \cap \P_v$) and a flux  $\cF_2$ on the second brane (at $p_2=\P_v\cap \P_\sigma^-$). According to the rules above, the first flux is lifted to $G_{4,1}\sim \cF_1\wedge  R^+_{\sigma}$, while the second one is lifted to $G_{4,1}\sim \cF_2\wedge  (R^+_{\sigma}+R_v)$. Imposing 4d Poincar\'e invariance and summing both contributions one gets
\begin{equation}
 G_4 = \frac12\left( \cF_1 + \cF_2  \right) \wedge \left( R^+_{\sigma} - R^-_{\sigma}\right) + \left( \cF_2 - \cF_1\right) \wedge R_v \:.
\end{equation}
We are interested in a flux along the Cartan generator of ${\rm SU}(2)$. We achieve this by choosing $\cF \equiv -\cF_1=\cF_2$. The corresponding `Cartan $G_4$-flux' is
\begin{equation}\label{G4cartanLocal}
G_4^{\rm Cartan} = \cF \wedge R_v  = \cF \wedge E \:.
\end{equation}
where we have used the fact that $R_v$ is actually the exceptional divisor $E=\{v=0\}$ in $\tilde{W}_E$.  This form was previously known to give the right lift of a Cartan flux to a $G_4$-flux on an elliptic fibration. In fact, the flux \eqref{G4cartanLocal} is well defined also in a generic elliptic fourfold of the family $W_5$:
\begin{equation}
 W_5\,: \qquad  \left\{ \begin{array}{l} (y-s\,z)(y+s\,z) = v\,s^3 + \sigma\, s\, z^2(v\,s +2\epsilon\,z^2)  \\ \sigma\,v=x_3 \\ \end{array} \right. \:.
\end{equation}
In \cite{Braun:2011zm} we verified that the global $Sp(1)$ realization of \eqref{G4cartanLocal} reproduces the same D3-charge and chiral modes as the Cartan two-form flux $\cF$.

\section{Hypercharge flux}\label{sec:HyperchFl}

In the previous sections, we have seen how a two-form flux on a D7-brane is lifted to a four-form flux in F-theory. We now want to apply this machinery to a flux particularly important for ${\rm SU}(5)$ GUT model building. 

In heterotic string compactifications, the St\"uckelberg mechanism unavoidably generates a mass for the ${\rm U}(1)_Y$ gauge field.
In contrast, in type IIB string theory, the GUT group can be broken to the Standard Model group by switching on a particular two-form flux along the hypercharge Cartan generator of ${\rm SU}(5)$ in F-theory \cite{Beasley:2008kw,Donagi:2008kj}, \emph{without} making it massive. Here, the GUT group can be realized on a stack of D7-branes wrapping a hypersurface $S$ of the Calabi-Yau threefold. The D7-brane worldvolume action has a coupling between the RR field $C_4$ and the gauge field strength
\begin{equation}\label{CSD7term}
\int_{D7} C_4\wedge \mbox{Tr} \,\hat{F}\wedge \hat{F} \:.
\end{equation}
Expanding the eight-dimensional field strength into an external (along $\mathbb{R}^{1,3}$) and an internal part (along $S$) as $\hat{F}=F+\cF$ and proceeding with the dimensional reduction of \eqref{CSD7term}, gives the 4D axionic coupling
\begin{equation}
  S_{Stk}=  \sum_{\ell,\alpha}K^\ell_\alpha  \int_{\mathbb{R}^{1,3}} F^\ell \wedge c_2^\alpha  \qquad\mbox{with}\qquad K^\ell_\alpha =  \mbox{Tr}(t^\ell)^2  \int_S \cF^\ell\wedge \imath^\ast \omega_\alpha   \:.
\end{equation}
Here $\{\omega_\alpha\}$ is basis of $H^{1,1}_+ (X_3)$, $\imath^\ast \omega_\alpha$ their pull-back to $S  \lhook\joinrel\xrightarrow{\imath} X_3$, $\ell$ runs over the (single) D7-branes and $c_2^\alpha$ are two-forms arising from the expansion of the RR four-form potential $C_4$ along the basis $\{\omega_\alpha\}$. 
When $K^\ell_\alpha\neq 0$ (for some $\alpha$) $A_\mu^\ell$ gets a mass from a St\"uckelberg mechanism. $A_\mu^Y$ is massless when $\int_S \cF^Y\wedge \imath^\ast\omega_\alpha =0$ $\forall \alpha$. Notice that this does not imply that the pushforward of the form $\cF^Y$ to the Calabi-Yau threefold is trivial, as pointed out in \cite{Buican:2006sn} and recently applied by \cite{Mayrhofer:2013ara} in the F-theory context. On the other hand, if the pushforward \emph{is} trivial in $X_3$, this is sufficient to have a massless ${\rm U}(1)_Y$ gauge field.
Since the odd two-forms of $X_3$ are projected out in the quotient $B_3$, the masslessness condition always requires $\iota_!\cF^Y=0$, where $\iota_!$ is the pushforward map from $S$ to $B_3$.

In the following, we present some examples of two-forms that satisfy the condition $\iota_!\cF=0$. We will use them to switch on a Cartan flux in type IIB that does not give mass to the associated generators. Then, we will lift such a two-form flux to $W_E$ by following the prescription outlined in the previous sections. The resulting four-form flux will be well defined also away from the weak coupling limit. This will give an explicit realization of a Cartan four-form flux that does not generate a St\"uckelberg mass for the associated Cartan gauge field.

\subsection{General case: $A_1$ singularity}\label{HypGenCase} \label{Sec:Sp1Glob}

We consider a compact fourfold and we enforce a split $A_1$ singularity along a surface $S$ given by $P=0$ in the base $B_3$. This is realized by the following factorizations:\footnote{Here (and in the following) $a_1$ is rescaled by a factor of $2$ with respect to the canonical Tate form.}
\begin{equation}
  b_2 \equiv a_1^2 + a_{2,1}\cdot P \,, \qquad\qquad  b_4 \equiv b_{4,1}\cdot P \,, \qquad\qquad  b_6 \equiv b_{6,2}\cdot P^2 \:.
\end{equation}
We resolve this $A_1$ singularity as in the local example. Then, the resolved fourfold $\tilde{W}_E$ is \footnote{This fourfold still has a conifold singularity which disappears
when going to the strong coupling elliptically fibered fourfold.}
\begin{equation}\label{Sp1resolSL}
\tilde{W}_E:\qquad \left\{ \begin{array}{l}
eq_E \equiv - (y-a_1s)(y+a_1s)+ \sigma(a_{2,1} s^2v+ 2 b_{4,1} s t  + b_{6,2} t^2\sigma )\\
\sigma \, v = P \\
\end{array} \right. 
\end{equation}
in an ambient sixfold, determined by the weight system
\begin{center} 
\begin{tabular}{ccccc}
$y$ & $s$ & $t$ & $\sigma$ & $v$ \\
$1$ & $1$ & $1$ & $0$ & $0$ \\
$1$ & $1$ & $0$ & $1$&$-1$ \\
$3\bar{K}$ & $2\bar{K}$ & $0$ & $[P]$ & $0$ 
\end{tabular}\\
\end{center}
and SR-ideal $[yst],[ys\varsigma],[vt]$.

Let us now choose the polynomial $P$ to take the following factorized form: 
\begin{equation}
 P \equiv A_a A_b -C D \:.
\end{equation}
For simplicity, we also choose the classes $[A_a]=[A_b]=[P]/2$, but the following results hold for generic $[A_a]$, $[A_b]$ (we just need that $q_a [A_a]= q_b [A_b]$ for some integers $q_a,q_b$).

The following two-cycles of $B_3$ are inside the D7-brane at $P=0$:
\begin{equation}\label{Cab2cyc}
 \cC_{a,b}\,: \qquad \{A_{a,b}=0 \} \qquad \cap \qquad \{ C=0\} \:.
\end{equation}
Moreover they are homologous in $B_3$ but non-homologous on the surface $P=0$. 
In the double cover Calabi-Yau threefold $X_3$ given by the equation $\xi^2=a_1^2 + a_{2,1}P$, the double cover of the surface $S$ splits into two surfaces $S^\pm$ isomorphic to $S$ itself. They are given by the equations $\{\xi\pm a_1=0\}\cap\{P=0\}$, which are exchanged by the involution.\footnote{We make the assumption of a smooth space $X_3$, i.e. we assume that the locus $\xi=a_1=P=a_{2,1}=0$ is empty \cite{Mayrhofer:2013ara}.}
Let us consider the surface $S^+$. The two cycles $\cC_{a,b}^+$ given by $\xi+a_1=A_{a,b}=C=0$ are homologous on $X_3$ but non-homologous on $S^+$.

We now construct a two form that is trivial in $X_3$ but non-trivial on the D7-worldvolume $S^+$. Note that here we are using the stronger masslessness constraint. 
This `trivial' two-form flux is given in terms of the Poincar\'e dual (in $S^+$) two-cycles as
\begin{equation}\label{HypF2Toy}
 F^+ = \cC_a^+ - \cC_b^+ \:.
\end{equation}
It projects down to a two-form  $F=\cC_a-\cC_b$ on $S\subset B_3$, whose pushforward to $B_3$ is trivial.
This is a toy model of hypercharge flux.

Let us now lift $F$ to the fourfold $\tilde{W}_E$:
\begin{equation}\label{Sp1resolSL2}
\tilde{W}_E:\qquad \left\{ \begin{array}{l}
eq_{E} =0\\
\sigma \, v = A_a A_b -C D \\
\end{array} \right. 
\end{equation}
The gauge flux is along the Cartan generator in type IIB, so that $G_4$ is obtained by lifting the flux $F$ via $\tilde{R}_v$. 
The resulting $G_4$-flux is given by fibering the exceptional $\P^1$ over the curves $\cC_{a,b}$ and taking the difference of the resulting fourcycles:
\begin{equation}
 G_4=\gamma_a-\gamma_b \qquad\mbox{where}\qquad \gamma_{a,b}\,: \,\, \{ eq_E =0\,, \,\,\, v= 0\,,\,\,\, A_{a,b}=0\,,\,\,\, C=0\}
\end{equation}
This 4-form flux is trivial in the ambient sixfold, but it is non-trivial on the Calabi-Yau fourfold. Note that, contrary to the Cartan $G_4$-fluxes constructed above, it is not a wedge product of the exceptional divisor with a two-form in $B_3$. However, it is Poincar\'e dual to a $\P^1$ fibration over curves on the brane locus. This form for the hypercharge flux was anticipated in \cite{Mayrhofer:2013ara}. Here we provide a recipe for an explicit realization.

As a check of this construction, one can show that the D3-charges of $F_2$ and $G_4$ are the same.

\subsection{Example: a rigid $\P^1\times \P^1$ in $B_3$}

In realistic models, one prefers to have the GUT brane wrapped on a rigid divisor in order to avoid possible exotics. This requirement seems to conflict with the construction we have given above: in fact, generically deformations of the polynomial $P$ are related to deformations of the D7-brane locus. We now give simple example in which the visible sector wraps a rigid divisor and we can still switch on Cartan flux without giving a (St\"uckelberg) mass to the corresponding gauge field.

\subsubsection*{A rigid surface with a two-cycle trivial in the base}\label{Sec:geometryToy}

We construct the base $B_3$, 
by blowing up $\P^4$ in a point, so that we get a toric variety $Y_4$ with the weight system
\begin{equation}
\begin{array}{cccccc}
 z_1 & z_2 & z_3 & z_4 & z_5 & w \\
 1 & 1& 1&1&1 & 0\\
 0&0&0&0&1&1 
\end{array}
\end{equation}
$Y_4$ has two independent divisor classes, which we can take as $[z_1]\equiv[z]$ and $[w]$. The exceptional locus $w=0$ is a $\P^3$. From the SR-ideal it follows that 
$[z]^4=0$ and $([w]+[z])[w]=0$. The intersection ring of $Y_4$ is hence
\begin{equation}\label{intringX}
\begin{array}{ccccc}
 [z]^4 & [z]^3[w] & [z]^2[w]^2 & [z][w]^3 & [w]^4 \\
 0 & 1 & -1 & 1 & -1 
\end{array}
\end{equation}
We take the base manifold $B_3$ as a hypersurface in $Y_4$ which is equivalent to the class $2[z_1]+[z_5]=3[z]+[w]$. More concretely, this has the form
\begin{equation}
B_3:\quad\quad P_2(z_1,\cdots,z_4)\,z_5 + w \, P_{3}(z_1,\cdots,z_4)=0\, .
\end{equation}

One can show that $3[z]+[w]$ defines an ample bundle on $Y_4$, so that the Lefschetz hyperplane theorem gives $b_2(B_3)=b_2(Y_4)=2$. \footnote{The Segre embedding shows that products, and more generally fiberd products of projective spaces, are projective. For projective varieties, we can use the Nakai-Moishezon criterion for ampleness. We compute $ \int_X c_1(N_{B_3})^4= \left(3[z]+[w]\right)^4 =65 >0$, so that the line bundle associated to $[B_3]$ is indeed ample and we can use the Lefschetz-hyperplane theorem. As we are talking about a line bundle on a toric variety, on can alternatively compute the piecewise linear support function corresponding to the divisor $3[z]+[w]$ and check strong convexity, from which ampleness follows \cite{Fulton}.}

We can compute the first Chern class of the base manifold:
\begin{equation}
 c_1(B_3) = c_1(Y_4) - c_1(N_{B_3})= 2[z]+[w] \:.
\end{equation}

The intersection ring on $B_3$ is generated by intersection of $[B_3]$ with $[w]$ and $[z]$. Using \eqref{intringX}, it is given by 
\begin{equation}\label{intringB3}
\begin{array}{cccc}
 [z]^3 & [z]^2[w] & [z][w]^2 & [w]^3  \\
 1 & 2 & -2 & 2  
\end{array}
\end{equation}

We now choose $S=B_3\cdot[w]$. This is a quadric in $\P^3$, so that $S \cong \P^1\times \P^1$ and we can choose coordinates such that
\begin{equation}
P_2=z_1 z_4-z_2 z_3=0 \:.
\end{equation} 
The surface $S$ is rigid in $B_3$: there is no deformation of $S$ which does not deform $B_3$ at the same time. The explicit identification of $S$ with $\P^1\times \P^1$ is made by using the homogeneous coordinates $[s_0:s_1]$ and $[t_0:t_1]$ on $\P^1\times \P^1$ and identifying
\begin{equation}
 (z_1,z_2,z_3,z_4) \leftrightarrow (s_0t_0,s_0t_1,s_1t_0,s_1t_1) \, .
\end{equation}
Hence $H_2(S,\mathbb{Z})$ is generated by the two $\P^1$'s
\begin{equation}\label{C12C13}
\cC_{12} = \{ x_1=x_2=w=0\} = \{s_0=0\} \qquad \mbox{and} \qquad \cC_{13}=\{ x_1=x_3=w=0\} = \{t_0=0\}\:,
\end{equation}
Furthermore, $[\cC_{12}]=[\cC_{13}]$ in $Y_4$, so that the Lefschetz hyperplane theorem implies that the cycle $\cC_{12}-\cC_{13}$ is trivial in $B_3$ even though it is non-trivial on $S$. Using the fact that $S$ is the product of the two rational curves $\cC_{13}$ and $\cC_{13}$, it is also clear that $ \int_S (\cC_{12}-\cC_{13})^2 = -2 $.

\subsubsection*{Toy GUT model with hypercharge flux and F-theory lift}

We can now use this geometry to build a toy model. The surface $S$ at $w=0$ will be the GUT divisor in $B_3$. For simplicity we will choose the simplest non-abelian group, i.e. ${\rm SU}(2)$, as the `GUT group'. 
Hence the fourfold $X_4$ must develop an $A_1$ singularity along the locus $w=0$:
\begin{equation}
 W_E\,: \qquad \left\{\begin{array}{l}
 y^2 = s^3 + (a_1^2+a_{2,1}w) s^2 + 2 b_{4,1}w\, s\,t + b_{6,2}w^2\, t^2 \\ \\
  (z_1 z_4 - z_2 z_3)\,z_5 + w  P_{3}(z_1,\cdots,z_4)=0
 \end{array}\right.
\end{equation}
where we used the form $P_2(z_1,...,z_4)=z_1 z_4 - z_2 z_3$.

We now consider a stack of two D7-branes wrapping the surface $S=\{w=0\}$ in $B_3$. 
Since $S$ is rigid and has $h^{0,1}=0$, there are no `exotic' scalars in the adjoint of the GUT group. On this brane we switch on a Cartan flux 
\begin{equation}
  F_2 = (\cC_{12}-\cC_{13})\otimes t^C
\end{equation}
where $\cC_{12}$ and $\cC_{13}$ are described in \eqref{C12C13} and $t^C$ is the Cartan generator. 
The two-cycle Poincar\'e dual to $F_2$ is non-trivial on the brane worldvolume, but it is trivial in $B_3$ (and on the double cover $X_3$), so that the ${\rm U}(1)$ gauge boson associated to the Cartan generator remains massless. Moreover, this flux does not generate chiral matter on the bulk of the branes.\footnote{Since $\int_S F_2^2=-2$ in the quotient surface $S=\P^1\times\P^1$, there is also no vector-like matter from the bulk (see \cite{Donagi:2008ca,Beasley:2008kw}).} 

Let us now consider the resolution of the singular F-theory fourfold: 
\begin{equation}
 \tilde{W}_E\,: \qquad \left\{\begin{array}{l}
 y^2 = s^3v + (a_1^2 + a_{2,1} w\,v)s^2 + 2 b_{4,1}w\, s\,t + b_{6,2}w^2\, t^2 \\ \\
  (z_1 z_4 - z_2 z_3)\,z_5 + v\,w\, P_{3}(z_1,\cdots,z_4)=0
 \end{array}\right.
\end{equation}
We construct the lift of the hypercharge flux to F-theory by following the procedure outlined in the sections \ref{Sec:NnAbSU2loc} and \ref{HypGenCase}. 
The resulting four-form flux is
\begin{equation}
 G_4^{Y} = \theta_{12} - \theta_{13}
\end{equation}
with
\begin{equation}\label{thetaG4YP1P1}
 \theta_{12}\,: \,\, \left\{\begin{array}{l}
 y^2 = a_1^2 s^2 + 2 b_{4,1}w\, s\,t + b_{6,2}w^2\, t^2 \\
  z_1 =0 \\
  z_2=0 \\ 
  v=0
 \end{array}\right.
\qquad
 \theta_{13}\,: \,\, \left\{\begin{array}{l}
 y^2 = a_1^2 s^2 + 2 b_{4,1}w\, s\,t + b_{6,2}w^2\, t^2 \\
  z_1 =0 \\
  z_3=0 \\ 
  v=0
 \end{array}\right.
\end{equation}

As expected, $G^Y_4$ is trivial in the ambient space but non-trivial on the fourfold $W_E$. 
Furthermore, this flux survives away from weak coupling. Its definition is a slight deformation of the equations \eqref{thetaG4YP1P1}, where the conic bundle is substituted by the (resolved) Weierstrass model, and the other equations remain untouched. 

This construction can be repeated for more realistic gauge groups, as we will see in the model presented in the next section.

\section{GUT on a $dP_7$ in $B_3$}

Following the prescription given in the last section, we can now construct a hypercharge flux directly in an F-theory ${\rm SU}(5)$ GUT model, without passing through the weak coupling limit. 

We consider an elliptic fourfolds over the base $B_3$ which is a hypersurface in the toric ambient space $Y_4$ with weight system \cite{Collinucci:2009uh} \footnote{We choose this base manifold because it gives a weak coupling limit with a smooth $X_3$ also when $b_2$ has the form for split ${\rm SU}(n)$ singularities \cite{Krause:2012yh}. We have then been able to check the F-theory results by type IIB computations at weak coupling.}
\begin{equation}\label{AmbY4}
\begin{array}{cccccc}
 z_1 & z_2 & z_3 & z_4 & z_5 & z_6 \\
 1 & 1& 1& 2& 0& 1\\
 0&0&0&1&1&1 
\end{array}
\end{equation}
and its SR-ideal is generated by $[z_1z_2z_3] , [z_4z_5z_6]$.
The base manifold $B_3$ is defined as the vanishing locus of a polynomial of degree $(5,2)$ in $Y_4$:
\begin{equation}\label{B3dP7md}
 B_3\,: \qquad P^{(5,2)}(z_1,...,z_6)=0
\end{equation}
Its first Chern class is $c_1(B_3) = \imath^\ast[z_6]$. The divisor $z_5=0$ is a dP$_7$.

\subsection{${\rm SU}(5)\times {\rm U}(1)$ model}

We construct a model with an ${\rm SU}(5)$ gauge group and one massless ${\rm U}(1)$. In F-theory, the first requirement is satisfied by enforcing singular fibers of type $I_5$ over a divisor in the base, while the second one is realized by taking $a_6\equiv 0$ \cite{Grimm:2010ez}. This boils down to the following restrictions on the Tate form coefficients:
\begin{equation}
 a_2 \equiv z_5\cdot a_{2,1} \qquad a_3 \equiv z_5^2\cdot a_{3,2} \qquad a_4 \equiv z_5^3\cdot a_{4,3} \qquad a_6 \equiv 0
\end{equation}
where we have chosen $z_5=0$ as the ${\rm SU}(5)$ locus on $B_3$. The resulting fourfold has a split ${\rm SU}(5)$ singularity along $z_5=0$ and a conifold singularity along $a_{3,2}=a_{4,3}=0$.

The resolved fourfold is given by two equations in a six-dimensional toric ambient space $Y_6$ with the weight system
\begin{equation}\label{ResAmbSp}
\begin{array}{cccccccccrrrrr}
 z_1 & z_2 & z_3 & z_4 & z_5 & z_6 & X & Y & Z & v_1 & v_2 & v_3 & v_4 & \ell \\
 1 & 1& 1& 2& 0& 1& 2& 3& 0& 0& 0& 0& 0& 0\\
 0 & 0& 0& 1& 1& 1& 2& 3& 0& 0& 0& 0& 0& 0\\
 0 & 0& 0& 0& 0& 0& 2& 3& 1& 0& 0& 0& 0& 0\\
 0 & 0& 0& 0& 1& 0& 1& 1& 0& -1& 0& 0& 0& 0\\
 0 & 0& 0& 0& 0& 0& 1& 0& 0& 0& -1& 0& 1& 0\\
 0 & 0& 0& 0& 0& 0& 0& 1& 0& 0& 1& -1& 0& 0\\
 0 & 0& 0& 0& 0& 0& 0& 1& 0& 1& 0& 0& -1& 0\\
 0 & 0& 0& 0& 0& 0& 1& 1& 0& 0& 0& 0& 0& -1\\
\end{array}
\end{equation}
and SR-ideal is generated by $
[XY],\, [Z\ell],\, [z_5\ell],\, [v_1\ell],\, [v_4\ell],\, [v_1Z],\, [v_2Z],\, [v_3Z],\, [v_4Z],\, [v_1Y]$,
$[v_2Y],\, [v_3X],\, [v_4X],\, [z_5v_2],\, [z_5v_3],\, [v_1v_3]$.\\
\noindent
The equations defining the resolved fourfold $\tilde{X}_4$ are:
\begin{equation}
\left\{
\begin{array}{l}
 v_3v_4\ell\,Y^2+a_1\ell\,X\,Y\,Z+v_1v_4a_{3,2}z_5^2Y\,Z^3 = v_1v_2^2v_3\ell^2X^3+v_1v_2a_{2,1}z_5\ell\,X^2Z^2+v_1^2v_2v_4a_{4,3}z_5^3X\,Z^4 \\   \\
 z_5v_1v_2v_3v_4 \left(z_5v_1v_2v_3v_4Q_5+z_4 R_3+z_6R_4 \right) + \left[ z_4^2 P_1+z_4z_6P_2+z_6^2P_3 \right] = 0
\end{array}\right.
\end{equation}
where $Q_i,R_i,P_i$ are polynomials of degree $i$ in the coordinates $z_1,z_2,z_3$. We will schematically refer to these two equations as
\begin{equation}
 \tilde{X}_4\,: \, 
\left\{
\begin{array}{l}
 eq_1  = 0  \\
 eq_2  = 0
\end{array}\right.
\end{equation}

\begin{figure}[!h]
\centering
\scalebox{0.4}{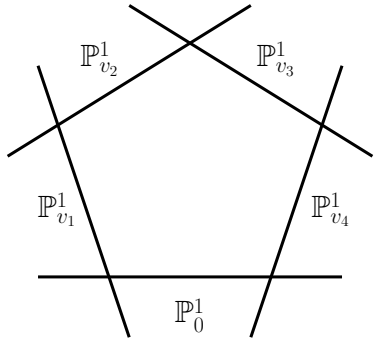}
\caption{The fiber components of a fiber of type $I_5$ meet according to the extended Dynkin diagram of ${\rm SU}(5)$. Here we have drawn each $\P^1$ as a line.
\label{I5}}
\end{figure}

There are Cartan divisors $E_i$ related to the vanishing of the four coordinates $v_i$, which are $\mathbb{P}^1$ fibrations over 
the surface $z_5=0$ on the base $B_3$. On top of $z_5=0$, there are five fiber components (including components coming from the divisors $E_i$)
which intersect as the extended Dynkin diagram $A_4$, see figure \ref{I5}.

\subsection{Hypercharge $G_4$-flux}

Let us now proceed in analogy with the ${\rm SU}(2)$ case and construct the hypercharge $G_4$-flux.
To obtain a Cartan flux that does not produce a massive ${\rm U}(1)$, we first need to choose a curve 
on the surface $S=\{z_5=0\}$ on the base $B_3$ that is trivial on $B_3$ itself. This surface is a dP$_7$ described by the equation 
\begin{equation}
 eq_S \equiv \left[ z_4^2 P_1(z_1,z_2,z_3)+z_4z_6P_2(z_1,z_2,z_3)+z_6^2P_3(z_1,z_2,z_3) \right] = 0
\end{equation}
in the toric ambient space\footnote{Note that the ambient space of the dP$_7$ is simply a blow-up of $\P^3$. In fact this dP$_7$ is 
the blow-up of a dP$_6$ which is given as a cubic in $\P^3$.}
\begin{equation}
\begin{array}{ccccc}
 z_1 & z_2 & z_3 & z_4 & z_6 \\
 1 & 1& 1& 2& 1\\
 0&0&0&1&1 
\end{array} \qquad \mbox{or equivalently} \qquad 
\begin{array}{ccccc}
 z_1 & z_2 & z_3 & z_4 & z_6 \\
 1 & 1& 1& 1& 0\\
 0&0&0&1&1 
\end{array}
\end{equation}
with SR-ideal generated by $[z_1z_2z_3]$ and $[z_4z_6]$.

In order to make new holomorphic curves appear as algebraic cycles (analogously with the case $\P^1\times \P^1$), we 
need to appropriately restrict the defining equation. We make the Ansatz,
\begin{equation}
 eq_S^r = z_1z_2z_6 F_1(z_1z_6,z_2z_6,z_3z_6,z_4)+z_3z_4\tilde{F}_1(z_1z_6,z_2z_6,z_3z_6,z_4)\, ,
\end{equation}
where $F_1,\tilde{F}_1$ are linear combinations of their arguments. This gives rise to the following three independent algebraic curves on $S$:
\begin{equation}
 \begin{array}{llcl}
  C_{13}\,: & \{z_1=0\} & \cap & \{z_3=0\} \\
  C_{63}\,: & \{z_6=0\} & \cap & \{z_3=0\} \\
  C_{24}\,: & \{z_2=0\} & \cap & \{z_4=0\} 
 \end{array}
\end{equation}
On $B_3$ these three curves satisfy the linear relation:
\begin{equation}\label{relcurves}
  C_{13} + C_{63} = C_{24} \:.
\end{equation}
We note that the curve $C_{63}$ can be written as a complete intersection of $z_6=0$ with $eq_S^r=0$. This is in contrast to $C_{13}$ and $C_{24}$, which 
can only be given by two equations in the ambient space that automatically satisfy $eq_S^r=0$. Their intersection numbers in $S$ are:
\begin{equation}
 C_{13}^2=-2\,\qquad C_{63}^2=-1\,\qquad C_{24}^2=-1\,\qquad C_{13}\cdot C_{63} =1\,\qquad C_{13}\cdot C_{24} =0\,\qquad C_{63}\cdot C_{24} =0 \:.
\end{equation}
From these numbers, one finds\footnote{Note that \eqref{relcurves} only holds on $B_3$, but does not say that $C_{13} + C_{63} - C_{24}$ is also trivial on $S$.} 
$(C_{13} + C_{63} - C_{24})^2=-2$. 

The hypercharge flux is constructed by lifting each of the curves to four-cycles on $\tilde{X}_4$. These four-cycles are combinations of $\P^1$ fibrations over the given curves, where the $\P^1$ are the exceptional 2-spheres on top of $v_i=0$. For ${\rm SU}(2)$ there was only $\P^1$ in the fiber, now we can construct the $\P^1$-fibred fourcycles 
\begin{eqnarray}\label{thetaiForYfl}
 \theta^{i}_{13} &:& v_i=0 \qquad z_1=0 \qquad z_3=0  \qquad eq_1=0 \qquad \subset\, Y_6 \nonumber \\
 \theta^{i}_{63} &:& v_i=0 \qquad z_6=0 \qquad z_3=0  \qquad eq_1=0 \qquad \subset\, Y_6 \\
 \theta^{i}_{24} &:& v_i=0 \qquad z_2=0 \qquad z_4=0  \qquad eq_1=0 \qquad \subset\, Y_6 \nonumber
\end{eqnarray}
associated with the curves $C_{13},C_{63},C_{24}$. The combination associated with the hypercharge flux must be the one related to the hypercharge generator, i.e.
\begin{equation}
  T^Y = \left(
\begin{array}{ccccc}
 -2 &&&& \\ & -2 &&& \\ && -2 && \\ &&& \hspace{1ex} 3 & \\ &&&& \hspace{1ex} 3 \\
\end{array}\right) = 2 T^{\alpha_1} + 4 T^{\alpha_2} + 6 T^{\alpha_3} +3T^{\alpha_4}  \:,
\end{equation}
where $\alpha_i$ are the simple roots of the $A_4$ Lie algebra.

Hence we can form a fourcycle along the hypercharge direction for every curve $C$ by
\begin{equation}
  \theta^Y_C = -2 \theta^1_C - 4 \theta^2_C - 6 \theta^3_C - 3 \theta^4_C  \:.
\end{equation}
A hypercharge flux $G_4^Y$ trivial in the ambient space can then be constructed as
\begin{equation}
  G_4^Y = \theta^Y_{24} - \theta^Y_{13} - \theta^Y_{63}  \:
\end{equation}
in terms of its Poincar\'e dual (and non-factorizable) 4-cycle in $\tilde{X}_4$. 
Due to \eqref{relcurves} this 4-cycle is trivial as a four-cycle in the ambient space $Y_6$, but is non-trivial in $\tilde{X}_4$.

In the following we see how to use such an object in F-theory model building.

\subsection{Matter curves and matter surfaces}

Matter fields in the ${\bf 10}$ and ${\bf 5}$ representations of ${\rm SU}(5)$ are localized on loci of $B_3$ where the $A_4$ singularity is enhanced to $D_5$ or $A_5$, respectively. 
In the ${\rm SU}(5)\times {\rm U}(1)$ model, the ${\bf 5}$ curve factorizes, while the ${\bf 10}$ curve does not. Summarizing, we have:
\begin{equation}
 \begin{array}{ccll}
  {\bf 10}_M & \leftrightarrow &  C_{{\bf 10}_M} \,:\,\, z_5 = a_1 =0 & \subset B_3 \\
  {\bf 5}_M & \leftrightarrow & C_{{\bf 5}_M} \,:\,\, z_5 = a_{3,2} = 0 & \subset B_3 \\
  {\bf 5}_H & \leftrightarrow & C_{{\bf 5}_H} \,:\,\, z_5 = a_1a_{4,3}-a_{2,1}a_{3,2} = 0 & \subset B_3 \\
 \end{array}
\end{equation}
On top of these surfaces, one of the exceptional $\P^1$'s splits, generating the structure of the extended Dynkin diagram of $D_5$ or $A_5$. Fibering the new $\P^1$'s over the matter curves gives `matter surfaces'.\footnote{The integral of the $G_4$-flux over these surfaces determine the number of chiral zero modes in the corresponding representations.} The way fiber components split over various loci in the base for the present model can be found in \cite{Krause:2011xj}. Here, we review just the ${\bf 5}$ curves:
\begin{itemize}
 \item ${\bf 5}_M$: $\P^1_{v_2}$ splits into $\P^1_{v_2,\ell}+\P^1_{v_2,E}$. Fibering $\P^1_{v_2,\ell}$ over the curve $C_{{\bf 5}_M}$ gives the associated matter surface: 
\begin{equation}
   \hat{\Sigma}_{{\bf 5}_M}\,: \qquad v_2=\ell=a_{3,2}=eq_2=0 \qquad \subset\, Y_6
\end{equation}
 \item ${\bf 5}_H$: $\P^1_{v_3}$ splits into $\P^1_{v_3,G}+\P^1_{v_3,H}$. Fibering $\P^1_{v_3,H}$ over the curve $C_{{\bf 5}_H}$ gives the associated matter surface:
\begin{equation}
   \hat{\Sigma}_{{\bf 5}_H}\,: \qquad v_3=a_1Y-v_1v_2a_{2,1}z_5XZ=a_1a_{4,3}-a_{2,1}a_{3,2}=eq_2=0 \qquad \subset\, Y_6
\end{equation}
\end{itemize}
These two surfaces only correspond to one of the five states in each of the ${\bf 5}$ representations of ${\rm SU}(5)$. The others can be constructed by adding or subtracting the surfaces 
\begin{equation}
   \left(\begin{array}{c} \Sigma_{{\bf 5}_M}^I \\ \Sigma_{{\bf 5}_M}^I \\ \end{array}\right) \,: \qquad v_I=
\left(\begin{array}{c}
a_{3,2} \\ a_1a_{4,3}-a_{2,1}a_{3,2} \\
\end{array}\right)
=eq_1=eq_2=0 \qquad \subset\, Y_6 \, ,
\end{equation}
where $I=0,1,...,4$ and $v_0\equiv z_5$. These correspond to the simple roots of the $A_4$ Dynkin diagram for $I=1,2,3,4$ and to the extra node for $I=0$. In particular \cite{Krause:2011xj}
\begin{equation}
 \vec{\Sigma}_{{\bf 5}_M} \, = \, \left( \begin{array}{c} 
    \Sigma_{{\bf 5}_M}^0+ \hat{\Sigma}_{{\bf 5}_M} + \Sigma_{{\bf 5}_M}^3 +\Sigma_{{\bf 5}_M}^4  \\ 
    \Sigma_{{\bf 5}_M}^0 + \Sigma_{{\bf 5}_M}^1+ \hat{\Sigma}_{{\bf 5}_M} + \Sigma_{{\bf 5}_M}^3 +\Sigma_{{\bf 5}_M}^4 \\ 
    \hat{\Sigma}_{{\bf 5}_M} \\ \hat{\Sigma}_{{\bf 5}_M} + \Sigma_{{\bf 5}_M}^3  \\  \hat{\Sigma}_{{\bf 5}_M} + \Sigma_{{\bf 5}_M}^3 +\Sigma_{{\bf 5}_M}^4 \\
 \end{array}\right) \qquad
\vec{\Sigma}_{{\bf 5}_H} \, = \, \left( \begin{array}{c} 
    \Sigma_{{\bf 5}_H}^0+ \hat{\Sigma}_{{\bf 5}_H} + \Sigma_{{\bf 5}_H}^4  \\ 
    \Sigma_{{\bf 5}_H}^0 + \Sigma_{{\bf 5}_H}^1+ \hat{\Sigma}_{{\bf 5}_H} + \Sigma_{{\bf 5}_H}^4 \\ 
    \Sigma_{{\bf 5}_H}^0 + \Sigma_{{\bf 5}_H}^1 + \Sigma_{{\bf 5}_H}^2 + \hat{\Sigma}_{{\bf 5}_H} + \Sigma_{{\bf 5}_H}^4 \\
        \hat{\Sigma}_{{\bf 5}_H} \\ 
 \hat{\Sigma}_{{\bf 5}_H} + \Sigma_{{\bf 5}_H}^4  \\  

 \end{array}\right)
\end{equation}
On can check that $G_4^Y$ integrates to zero over all these curves. This is good for the matter in the ${\bf 5}_M$ representation, as we want the same number of $({\bf 3},{\bf 1})$ and $({\bf 1},{\bf 2})$ after breaking ${\rm SU}(5)$ to the gauge group ${\rm SU}(3)\times {\rm SU}(2)$.
With regard to the Higgs, we would like to have a different number of doublets and triplets. In particular, it is desirable to have no triplet at all and only a non-chiral spectrum of doublets. This is possible if we split the Higgs matter curve $C_{{\bf 5}_H}$. 
In order to do this, we restrict the complex structure of the fourfold such that:
\begin{equation}
 a_{3,2} \equiv z_2 \hat{a}_{3,2}+a_1 Q_2(z_1,z_2,z_3) \qquad \mbox{ and } \qquad  a_{4,3} \equiv z_4 \hat{a}_{4,3}+ a_{2,1} Q_2(z_1,z_2,z_3) \:,
\end{equation}
with $Q_2(z_1,z_2,z_3)$ a generic polynomial of degree $2$ in the coordinates $z_1,z_2,z_3$. 
The equations defining the matter curve in the toric ambient space $Y_4$ (defined in \eqref{AmbY4}) becomes 
\begin{equation}
 C_{{\bf 5}_H} \,:\,\, z_5 = eq_S^r = z_4a_1\hat{a}_{4,3}-z_2a_{2,1}\hat{a}_{3,2} = 0 \:.
\end{equation}
Remember that $eq_S^r \equiv z_1z_2z_6 F_1(z_1z_6,z_2z_6,z_3z_6,z_4)+z_3z_4\tilde{F}_1(z_1z_6,z_2z_6,z_3z_6,z_4)$. Hence $C_{{\bf 5}_H}$ splits as
\begin{equation}
 C_{{\bf 5}_H} \rightarrow C_{{\bf 5}_{H_d}} + \, C_{{\bf 5}_{H_u}}  
\end{equation}
where 
\begin{equation}
 C_{{\bf 5}_{H_d}} \,:\,\, z_5 = z_2 = z_4 = 0 \,\,\, \subset \, Y_4 \qquad \mbox{and} \qquad C_{{\bf 5}_{H_u}} =C_{{\bf 5}_H} - C_{{\bf 5}_{H_d}}\:.
\end{equation}
Correspondingly, the matter surfaces split as well: $\vec{\Sigma}_{{\bf 5}_H} \rightarrow \vec{\Sigma}_{{\bf 5}_{H_d}}+\,\vec{\Sigma}_{{\bf 5}_{H_u}}$. Note that $ C_{{\bf 5}_{H_d}} $ is the same curve denoted by $C_{24}$ before. 

If we integrate $G_4^Y$ over these surfaces, we obtain
\begin{equation}\label{G4YSigmaH}
 \vec{\Sigma}_{{\bf 5}_{H_d}}\cdot G_4^Y = \left(\begin{array}{c} 2 \\ 2 \\ 2 \\ -3 \\  -3 \\ 
  \end{array}\right) \qquad\qquad
 \vec{\Sigma}_{{\bf 5}_{H_u}}\cdot G_4^Y = \left(\begin{array}{c} -2 \\ -2 \\ -2 \\ 3 \\  3 \\ 
  \end{array}\right)
\end{equation}
These intersections give the number of chiral states \cite{Donagi:2008ca,Hayashi:2008ba,Braun:2011zm,Krause:2011xj,Grimm:2011fx}.
As expected, we have a vector-like spectrum in the Higgs sector. As a further requirement,
we would also like the zero modes for the triplet to be absent (instead of having two). 
For this we need to switch on some different flux which gives the same number when 
integrated over all the matter surfaces related to one matter curve.

\subsection{Massless ${\rm U}(1)$ flux and doublet-triplet splitting}

The extra conifold singularity at $a_{3,2}=a_{4,2}=0$ (which comes from the condition
$a_6\equiv 0$) has been cured by a small resolution, which introduces the new coordinate $\ell$ 
in the last line in the table \eqref{ResAmbSp}. 
In the resolved space, the new section related to the extra ${\rm U}(1)$ is determined by the equation $\ell=0$ 
in the Calabi-Yau fourfold $\tilde{X}_4$. In an ${\rm SU}(5)\times {\rm U}(1)$ model, the new massless vector is 
produced by expanding the three-form $C_3$ along the the two-form \cite{Krause:2011xj}
\begin{equation}
 w_X = -5 ([\ell]-[Z]-c_1(B_3)) - 2E_1-4E_2-6E_3-3E_4\:,
\end{equation}
where $E_i$ are the exceptional divisors $v_i=0$. 
It is now easy to switch on the corresponding ${\rm U}(1)$ flux:
\begin{equation}
 G_4^X = F_X \wedge w_X
\end{equation}
where $F_X$ is a two-form along the base $B_3$. In the following we will make the choice
\begin{equation}
 F_X = 8 [z_6] - [z_1] \:,
\end{equation}
where we have written $F_X$ in terms of its Poincar\'e dual as a combination of divisors 
in $B_3$.\footnote{
The four-form fluxes we are considering, i.e. $G_4 = G_4^X + G_4^Y$ is an integral four-form. It satisfies the flux quantization condition if
$G_4+\frac{c_2(\tilde{X}_4)}{2} \in H^4(\tilde{X}_4,\mathbb{Z})$, so that we need to check if $c_2(\tilde{X}_4)$ is an even four-form. In the ${\rm SU}(5)\times {\rm U}(1)$ model we are studying, $c_2(\tilde{X}_4)$ is even up to the contribution $w_X\wedge ( [z_5]+\sum_{i=1}^4E_i)$. This four-form integrates to an even number over all our matter surfaces. On the other hand, it gives odd numbers when integrated over, for example, the surface $\theta_{13}^3$ in \eqref{thetaiForYfl}. This means that to produce a properly quantized flux, one needs to further add a 4-form flux $\delta G_4=\frac{w_X\wedge ( [z_5]+\sum_{i=1}^4E_i)}{2}-\Theta_4$ to $G_4^X+G_4^Y$. In order not to change the chirality, $\delta G_4$ must integrate to zero over all matter surfaces (in particular, $\Theta_4$ will be a non-factorizable four-form). We do not give details here, as this complicated computation will not add any relevant results. The existence of this flux can be established by studying the weak coupling limit along the lines 
of \cite{Krause:2012yh}.}

\

We can now compute the chiral number generated by this flux on the matter surfaces. The two-form $w_X$ is chosen such that it has zero intersection with the exceptional divisors $E_i$. Hence the number of intersection points of $G_4^X$ with all the matter surfaces corresponding to a given representation is the same. In our model we have 
\begin{equation}\label{G4XSigmaMH}
 \vec{\Sigma}_{{\bf 10}_{M}}\cdot G_4^X =  9 \qquad
 \vec{\Sigma}_{{\bf 5}_{M}}\cdot G_4^X = -9 \qquad
 \vec{\Sigma}_{{\bf 5}_{H_d}}\cdot G_4^X = -2 \qquad
 \vec{\Sigma}_{{\bf 5}_{H_u}}\cdot G_4^X = 2 
\end{equation}
where we have used $\vec{\Sigma}_{{\bf R}}\cdot G_4^X = q_{\bf R} \int_{C_{{\bf R}}} F_X$ \cite{Krause:2011xj}.
Since $G_4^Y$ does not intersect the matter surfaces $\vec{\Sigma}_{{\bf 10}_{M}}$ and $\vec{\Sigma}_{{\bf 5}_{M}}$, the number of matter chiral states is 
\begin{equation}
 n_{{\bf 10}_M} = 9 \qquad \qquad n_{{\bf \bar{5}}_M} = 9 
\end{equation}
Hence we have $9$ generations in this model.

This choice of $G_4^X$ also addresses the doublet-triplet splitting problem, i.e. the fact that in the MSSM we only have Higgs doublet and no Higgs triplet. In fact, by switching on $G_4^Y$, we break the gauge group from ${\rm SU}(5)$ to ${\rm SU}(3)\times {\rm SU}(2) \times {\rm U}(1)_Y$.  Accordingly the ${\bf 5}_{H_{d,u}}$ representations are broken as
\begin{equation}
 {\bf 5}_{H_{d,u}} \rightarrow ( {\bf 3}, {\bf 1} )^{-2_Y}_{H_{d,u}} + ( {\bf 1}, {\bf 2} )^{+3_Y}_{H_{d,u}} 
\end{equation}
By using \eqref{G4XSigmaMH} and \eqref{G4YSigmaH}, one can compute the corresponding chiral numbers for the triplet:
\begin{equation}
n_{ ( {\bf \bar{3}}, {\bf 1} )^{-2_Y}_{H_{d}}} = 0 \qquad n_{ ( {\bf 3}, {\bf 1} )^{+2_Y}_{H_{u}}} = 0 \qquad 
\end{equation}
On the other hand for the doublet we have a vector-like non-zero spectrum:
\begin{equation}
n_{ ( {\bf 1}, {\bf 2} )^{-3_Y}_{H_{d}}} = 5 \qquad n_{ ( {\bf 1}, {\bf 2} )^{3_Y}_{H_{u}}} = 5 \qquad 
\end{equation}

 We also see that the ${\rm SU}(5)$ spectrum is non-anomalous (as expected since the type IIB limit has zero D5-tadpole). One can also compute the number of singlets that live on the matter curve $C_{\bf 1}$ given by the equations $a_{3,2}=a_{4,3}=0$ in $B_3$. The corresponding matter surfaces $\Sigma_{\bf 1}$ is given by $\ell=a_{3,2}=a_{4,3}=eq_2=0$ in the ambient sixfold $Y_6$. The number of chiral states is 
\begin{equation}
n_{ {\bf 1}} = \int_{\Sigma_{\bf 1}} G_4^X + G_4^Y   = 5\int_{C_{\bf 1}} F_X = 1095 \:,
\end{equation}
where we used that $G_4^Y$ does not intersect the singlet matter surface.

\section{Lifting fluxes for $U(1)$ restriction and smooth Weierstrass model}\label{sec:u1resSmW}

In this section we will show how to reproduce the fluxes studied in \cite{Braun:2011zm,Grimm:2010ez} by the the lifting procedure outlined above. We will first study the case of brane/image-brane fluxes, where its application is straightforward. After this we treat the more general case for which the Weierstrass model is smooth. In both cases, \cite{Braun:2011zm} contained a conjecture for which two-form flux corresponds to a given four-form flux. This conjecture was supported by matching the D3-charges, the induced chiralities and the number of stabilized moduli. Here, we explicitly construct one flux starting from the other.

\subsection{Brane/Image-brane fluxes}


Let us start by again considering
\begin{align}
 W_E:\quad  &   y^2=b_2s^2 +2 b_4t + b_6t^2  \, .
\end{align}
For generic $b_2,b_4,b_6$, this $\P^1$ fibration does not have a section (the section $z=0$ of the Weierstrass model is contained in $W_T$). To generate a section, we can impose $b_6\equiv a_3^2$ \cite{Grimm:2010ez}. In this case the equation defining $W_E$ can be written as 
\begin{equation}
 (y-a_3 t)(y+ a_3 t)= s (b_2 s+2b_4 t)
\end{equation}
We notice the usual two things: 1) there are two new sections $\sigma_\pm$: $y\mp a_3 t = s=0$, and 2) the manifold has a conifold singularity along the curve $y=s=a_3=b_4=0$. One can make a small resolution to cure this singularity: after a change of coordinates $Y\equiv y-a_3t$ we set $(s,Y)\mapsto (\ell\,s,\ell\, Y)$, obtaining the scalings
\begin{center}
\begin{tabular}{ccccc}
$Y$ & $s$ & $t$ & $\ell$  \\
$1$ & $1$ & $1$ & $0$  \\
$1$ & $1$ & $0$ & $-1$
\end{tabular}\\
\end{center}
The equation defining the resolved fourfold is
\begin{equation}
 \hat{W}_E\,: \qquad Y ( \ell\,Y + 2 a_3t) = s (\ell\,b_2s+2b_4t)
\end{equation}
The two sections are given by the equations:
\begin{eqnarray}
\sigma_+ &:& \ell = eq_{\hat{W}_E} = 0 \\
\sigma_- &:& s = eq_{\hat{W}_E} = 0 
\end{eqnarray}
The choice $b_6=a_3^2$ produces the discriminant locus $\Delta_E=b_4^2-b_2a_3^2$. In the Calabi-Yau threefold $X_3$ embedded in $W_E$ ($t=0$), this locus splits into a brane and its image:
\begin{equation}
   \left\{\begin{array}{l}
   y^2=s^2b_2 \\ b_4^2-b_2a_3^2=0  \\
   \end{array}\right.  \qquad \leftrightarrow \qquad 
   \left\{\begin{array}{l}
   y^2=s^2b_2 \\ (s\,b_4-y\,a_3)(s\,b_4+y\,a_3)=0  \\
   \end{array}\right.  
\end{equation}
We want to lift a flux $\cF$ on such a brane (and $-\cF$ on its image).

The resolution outlined above also applies away from the central fiber in $W_5$. The generic resolved elliptic fibration has the form
\begin{equation}
Y(\ell \,Y +2\,a_3t\,z^3)= s (\lambda\,\ell^2 \,s^2 + \ell\,b_2s+2b_4t)\:.
\end{equation}
In \cite{Grimm:2010ez} (and in a different language in \cite{Braun:2011zm}) it was claimed that the two-form $w_{X_4}=-[\ell]+[z]+\bar{K}_B$ is related to the massless ${\rm U}(1)$ living on the brane $s\,b_4-y\,a_3=0$. The conjectured lift of $\cF$ was then
\begin{equation}\label{conjU1G4}
 G_4^{{\rm U}(1)} = \cF \wedge w_{X_4} \:.
\end{equation}
Away from the central fiber of the family $W_5$, this two-form is homologous to
\begin{equation}
 w_{X_4} = \frac12 \left( [s] - [\ell] \right) \:.
\end{equation}
In $W_E$, this two-form becomes
\begin{equation}
 w_{W_E} = \frac12 \left( \sigma_- - \sigma_+ \right) \:.
\end{equation}

Let's try to arrive at the result \eqref{conjU1G4} by applying the cylinder map. By a careful analysis, one realizes that $\hat{R}_3$ (the cylinder in the resolved space $\hat{W}_E$) splits into two pieces $\hat{R}_3=\hat{R}_3^+\cup \hat{R}_3^-$, with
\begin{equation}\label{R3pm}
\hat{R}_3^+ \,\,:  \,\, \hat{W}_E \,\, \cap \,\,  \left\{ \begin{array}{l}
b_2 a_3^2 - b_4^2=0 \\
s b_4 + a_3 Y  =0\\ 
s b_2 a_3 \pm b_4 Y =0\\
\end{array}\right. \qquad
\hat{R}_3^- \,\,:  \,\, \hat{W}_E \,\, \cap \,\,  \left\{ \begin{array}{l}
b_2 a_3^2 - b_4^2=0 \\
\ell\,s b_4 + a_3 (\ell\,Y+2a_3t)  =0\\ 
\ell\,s b_2 a_3 \pm b_4 (\ell\,Y+2a_3t) =0\\
\end{array}\right. 
\end{equation}
By applying the procedure that we outlined in the first non-compact model, one finds that the lift of $\cF$ is
\begin{equation}\label{G4U1wc}
 G_4^{{\rm U}(1),w.c.}= \frac12 \cF \wedge (\hat{R}_3^+ - \hat{R}_3^-) \:.
\end{equation}

As in the non-compact example, this flux is not well defined away from the central fiber in the family $W_5$. However one can prove the following relation in homology:\footnote{
Consider the following six-cycles, given by a complete intersection of $W_E$ with one equation:
\begin{equation}
\nu_\pm\,: \,\,\,  W_E  \qquad\cap\qquad s b_2 a_3 \pm b_4 (y \pm a_3t) =0\:. \nonumber
\end{equation}
We have that $[\nu_+]=[\nu_-]$ in $W_E$, so that $[\nu_+]-[\nu_-]$ is trivial.
Moreover, one can prove that $[\nu_\pm]$ splits as $[\nu_\pm]=2[\sigma_\pm]+[\hat{R}_3^\pm]$. Hence we have that $2[\sigma_+]+[ \hat{R}_3^+]=2[\sigma_-]+[ \hat{R}_3^-]$.}

\begin{equation}\label{equivRsigma}
 \hat{R}_3^+-\hat{R}_3^- = 2(\sigma_--\sigma_+)\:.
\end{equation}
Plugging this into \eqref{G4U1wc}, one obtains the conjectured expression \eqref{conjU1G4}.

\subsection{Non-factorizable flux}

Now we proceed to the case of a smooth fourfold $X_4$, for which the complex structure is restricted such that $b_6=a_3^2+\rho\tau$ \cite{Collinucci:2008pf}. In type IIB this corresponds to a recombination of the brane/image-brane system described in the last section. The resulting D7-brane is orientifold invariant and has the form of the Whitney umbrella \cite{Collinucci:2008pf}. In the threefold $X_3$, its worldvolume is spanned by the two points $t=0$ in the fiber
\begin{align}
 C_2\,: \qquad t=0 \qquad y^2=s^2b_2 \qquad (s \, b_4- y\, a_3)(s \, b_4- y\, a_3)  = s^2\rho\tau  \qquad \mbox{in} Y_5\, . 
\end{align}
One can define an odd two-form flux $F_2$, by taking the following two-cycles inside $C_2$,
\begin{equation}\label{2-cycDoubCov}
 \cC_\pm \, : \,\,\, t=0 \qquad y^2=s^2b_2 \qquad s b_4 = \pm  y a_3 \qquad \rho =0 \:.
\end{equation}
The flux is then
\begin{equation}\label{F2fluxW}
 F_2 = \frac12 ( \cC_+ - \cC_-) \:.
\end{equation}

In \cite{Braun:2011zm} we studied the non-perturbative limit of this configuration. A generic element $X_4$ of the family $W_5$ takes the form:
\begin{equation}
(y-a_3\,z^3)(y+a_3\,z^3)=s(s^2+b_2 s\,z^2 + 2b_4 z^4) + \rho\,\tau\,z^6 \:.
\end{equation}
We see that the restriction $b_6=a_3^2+\rho\tau$ on the complex structure makes the fourfold gain extra algebraic four-cycles. Typical examples of such cycles are
\begin{equation}\label{gammapmNP}
\gamma_{\rho\pm} \,: \qquad \{y\pm a_3 z^3=0\} \qquad \cap \qquad \{s=0\} \qquad \cap \qquad \{\rho=0\} \:.
\end{equation}
These are algebraic four-cycles in the ambient space $X_5$ which are completely sitting inside the Calabi-Yau fourfold $X_4$, but are not complete intersections of the Weierstrass equation with divisors of the ambient space. In \cite{Braun:2011zm}, we claimed that the two-form flux \eqref{F2fluxW} is equivalent to
\begin{equation}\label{PrevSub}
 G_4^\gamma = \gamma_{\rho+} - \frac12 [\rho]\cdot [s] \:.
\end{equation}
This flux can be written in a more covariant form: By intersecting $\{s=0\}$ and $\{\rho=0\}$ with the fourfold $X_4$, we note that $\gamma_{\rho+}+\gamma_{\rho-}=[s][\rho]$. Substituting this expression in \eqref{PrevSub} we obtain:
\begin{equation}\label{G4SubGamma}
 G_4^\gamma = \frac12\left( \gamma_{\rho+} - \gamma_{\rho-} \right) \:.
\end{equation}
In the weak coupling limit this flux keeps the same form in $W_E$, where now
\begin{equation}\label{gammapm}
\gamma_{\rho\pm} \,: \qquad \{y\pm a_3 t=0\} \qquad \cap \qquad \{s=0\} \qquad \cap \qquad \{\rho=0\} \:,
\end{equation}

We now derive this flux by applying the cylinder map to the two flux $F_2$ \eqref{F2fluxW}. 
Here, there is a new complication. As we have seen, the D7-brane on $X_3$ is an invariant brane wrapping the singular surface $C_2$. This is reflected in the cylinder $R_3=\Delta_E\cap W_E$, which develops a singularity of codimension one. As explained in \cite{Clingher:2012rg}, one needs to use the normalization of $C_2$ and $R_3$ in order to properly apply the cylinder map. After doing so, $R_3^{\rm norm}$ becomes a $\P^1$ fibration over $C_2^{\rm norm}$. One can then simply lift a curve $\cC$ in $C_2^{\rm norm}$ to a four-cycle $\theta$ in $R_3^{\rm norm}$ by fibering the $\P^1$ fiber over $\cC$. The four-cycle $\theta$ can then be uniquely mapped to a four cycle in $W_E$.

We start from the observation that the intersection of $R_3$ with the locus $\rho=0$ splits into two components of codimension two in $W_E$ (described by a set of non-independent equations):
\begin{equation}\label{thetapm}
\theta_\mp \,\,:  \qquad W_E \qquad \cap \qquad  \left\{ \begin{array}{l}
b_2 a_3^2 - b_4^2=0 \\
s b_4 \pm a_3 (y\pm a_3t) =0\\ 
s b_2 a_3 \pm b_4 (y \pm a_3t) =0\\
\rho=0 \\
\end{array}\right. 
\end{equation}
The four-cycle $\theta_+(\theta_-)$ 
is a $\P^1$ fibration over $\cC_+(\cC_-)$ with a section at $t=0$. 
It hence seems natural to use these cycles to lift the world-volume flux to F-theory. This is confirmed by applying the normalization procedure.

Following the definition of $F_2$, we construct the four-form flux by taking\footnote{The intersection of $\Delta_E=0$ with the locus $\rho=0$ splits into two components $\Delta_E^\pm$. The curve $\cC_\pm$ lies in the component $\Delta^\pm$. The orientifold involution exchanges the two components and $\cC_\pm$ as well. As done for the brane/image-brane system, we need to lift only one of these components to the physical space, or take the right combination with a factor of $1/2$. A further factor of $1/2$ comes from taking the flux $\cF_\pm=\frac12 \cC_\pm$.}
\begin{equation}
 G_4^\theta = \frac14 ( \theta_- - \theta_+)\:.
\end{equation}
By following the same steps we used to prove the equivalence \eqref{equivRsigma}, we find
\begin{equation}
 G_4^\theta = \frac14([\theta_-]-[\theta_+]) = \frac12([\gamma_{\rho +}]-[\gamma_{\rho -}])=G_4^\gamma \:.
\end{equation}
This proves the conjecture about the equivalence of $F_2$ and $G_4^\gamma$ 
made in \cite{Braun:2011zm}.

\section*{Acknowledgements}

We have benefited from discussions with Christoph Mayrhofer, Eran Palti, Raffaele Savelli, Taizan Watari and Timo Weigand.
A.C. is a Research Associate of the Fonds de la Recherche Scientifique F.N.R.S. (Belgium).
The work of A.P.B is supported by the STFC under grant ST/J002798/1.

\bibliography{hc.bib}

\providecommand{\href}[2]{#2}\begingroup\raggedright\begin{thebibliography}{10}

\bibitem{Buican:2006sn}
M.~Buican, D.~Malyshev, D.~R. Morrison, H.~Verlinde, and M.~Wijnholt,
  ``{D-branes at Singularities, Compactification, and Hypercharge},''
  \href{http://dx.doi.org/10.1088/1126-6708/2007/01/107}{{\em JHEP} {\bf 0701}
  (2007)  107},
\href{http://arxiv.org/abs/hep-th/0610007}{{\tt arXiv:hep-th/0610007
  [hep-th]}}.

\bibitem{Tatar:2008zj}
R.~Tatar and T.~Watari, ``{GUT Relations from String Theory
  Compactifications},''
  \href{http://dx.doi.org/10.1016/j.nuclphysb.2008.11.009}{{\em Nucl. Phys.}
  {\bf B810} (2009)  316--353},
\href{http://arxiv.org/abs/0806.0634}{{\tt arXiv:0806.0634 [hep-th]}}.

\bibitem{Blumenhagen:2008zz}
R.~Blumenhagen, V.~Braun, T.~W. Grimm, and T.~Weigand, ``{GUTs in Type IIB
  Orientifold Compactifications},''
  \href{http://dx.doi.org/10.1016/j.nuclphysb.2009.02.011}{{\em Nucl. Phys.}
  {\bf B815} (2009)  1--94},
\href{http://arxiv.org/abs/0811.2936}{{\tt arXiv:0811.2936 [hep-th]}}.

\bibitem{Beasley:2008dc}
C.~Beasley, J.~J. Heckman, and C.~Vafa, ``{GUTs and Exceptional Branes in
  F-theory - I},'' \href{http://dx.doi.org/10.1088/1126-6708/2009/01/058}{{\em
  JHEP} {\bf 01} (2009)  058},
\href{http://arxiv.org/abs/0802.3391}{{\tt arXiv:0802.3391 [hep-th]}}.

\bibitem{Beasley:2008kw}
C.~Beasley, J.~J. Heckman, and C.~Vafa, ``{GUTs and Exceptional Branes in
  F-theory - II: Experimental Predictions},''
  \href{http://dx.doi.org/10.1088/1126-6708/2009/01/059}{{\em JHEP} {\bf 01}
  (2009)  059},
\href{http://arxiv.org/abs/0806.0102}{{\tt arXiv:0806.0102 [hep-th]}}.

\bibitem{Donagi:2008ca}
R.~Donagi and M.~Wijnholt, ``{Model Building with F-Theory},''
\href{http://arxiv.org/abs/0802.2969}{{\tt arXiv:0802.2969 [hep-th]}}.

\bibitem{Donagi:2008kj}
R.~Donagi and M.~Wijnholt, ``{Breaking GUT Groups in F-Theory},''
\href{http://arxiv.org/abs/0808.2223}{{\tt arXiv:0808.2223 [hep-th]}}.

\bibitem{Marsano:2009ym}
J.~Marsano, N.~Saulina, and S.~Schafer-Nameki, ``{F-theory Compactifications
  for Supersymmetric GUTs},''
\href{http://arxiv.org/abs/0904.3932}{{\tt arXiv:0904.3932 [hep-th]}}.

\bibitem{Dudas:2010zb}
E.~Dudas and E.~Palti, ``{On hypercharge flux and exotics in F-theory GUTs},''
  \href{http://dx.doi.org/10.1007/JHEP09(2010)013}{{\em JHEP} {\bf 1009} (2010)
   013},
\href{http://arxiv.org/abs/1007.1297}{{\tt arXiv:1007.1297 [hep-ph]}}.

\bibitem{Marsano:2010sq}
J.~Marsano, ``{Hypercharge Flux, Exotics, and Anomaly Cancellation in F-theory
  GUTs},'' \href{http://dx.doi.org/10.1103/PhysRevLett.106.081601}{{\em
  Phys.Rev.Lett.} {\bf 106} (2011)  081601},
\href{http://arxiv.org/abs/1011.2212}{{\tt arXiv:1011.2212 [hep-th]}}.

\bibitem{Palti:2012dd}
E.~Palti, ``{A Note on Hypercharge Flux, Anomalies, and U(1)s in F-theory
  GUTs},'' \href{http://dx.doi.org/10.1103/PhysRevD.87.085036}{{\em Phys.Rev.}
  {\bf D87} (2013) no.~8, 085036},
\href{http://arxiv.org/abs/1209.4421}{{\tt arXiv:1209.4421 [hep-th]}}.

\bibitem{Mayrhofer:2013ara}
C.~Mayrhofer, E.~Palti, and T.~Weigand, ``{Hypercharge Flux in IIB and
  F-theory: Anomalies and Gauge Coupling Unification},''
  \href{http://dx.doi.org/10.1007/JHEP09(2013)082}{{\em JHEP} {\bf 1309} (2013)
   082},
\href{http://arxiv.org/abs/1303.3589}{{\tt arXiv:1303.3589 [hep-th]}}.

\bibitem{Weigand:2010wm}
T.~Weigand, ``{Lectures on F-theory compactifications and model building},''
  \href{http://dx.doi.org/10.1088/0264-9381/27/21/214004}{{\em
  Class.Quant.Grav.} {\bf 27} (2010)  214004},
\href{http://arxiv.org/abs/1009.3497}{{\tt arXiv:1009.3497 [hep-th]}}.

\bibitem{Maharana:2012tu}
A.~Maharana and E.~Palti, ``{Models of Particle Physics from Type IIB String
  Theory and F-theory: A Review},''
  \href{http://dx.doi.org/10.1142/S0217751X13300056}{{\em Int.J.Mod.Phys.} {\bf
  A28} (2013)  1330005},
\href{http://arxiv.org/abs/1212.0555}{{\tt arXiv:1212.0555 [hep-th]}}.

\bibitem{Denef:2007vg}
F.~Denef and G.~W. Moore, ``{Split states, entropy enigmas, holes and halos},''
  \href{http://dx.doi.org/10.1007/JHEP11(2011)129}{{\em JHEP} {\bf 1111} (2011)
   129},
\href{http://arxiv.org/abs/hep-th/0702146}{{\tt arXiv:hep-th/0702146
  [HEP-TH]}}.

\bibitem{Clingher:2012rg}
A.~Clingher, R.~Donagi, and M.~Wijnholt, ``{The Sen Limit},''
\href{http://arxiv.org/abs/1212.4505}{{\tt arXiv:1212.4505 [hep-th]}}.

\bibitem{Braun:2011zm}
A.~P. Braun, A.~Collinucci, and R.~Valandro, ``{G-flux in F-theory and
  algebraic cycles},''
  \href{http://dx.doi.org/10.1016/j.nuclphysb.2011.10.034}{{\em Nucl.Phys.}
  {\bf B856} (2012)  129--179},
\href{http://arxiv.org/abs/1107.5337}{{\tt arXiv:1107.5337 [hep-th]}}.

\bibitem{Grimm:2010ez}
T.~W. Grimm and T.~Weigand, ``{On Abelian Gauge Symmetries and Proton Decay in
  Global F-theory GUTs},''
  \href{http://dx.doi.org/10.1103/PhysRevD.82.086009}{{\em Phys.Rev.} {\bf D82}
  (2010)  086009},
\href{http://arxiv.org/abs/1006.0226}{{\tt arXiv:1006.0226 [hep-th]}}.

\bibitem{Donagi:2012ts}
R.~Donagi, S.~Katz, and M.~Wijnholt, ``{Weak Coupling, Degeneration and Log
  Calabi-Yau Spaces},''
\href{http://arxiv.org/abs/1212.0553}{{\tt arXiv:1212.0553 [hep-th]}}.

\bibitem{Gubser:2002mz}
S.~S. Gubser, ``{TASI lectures: Special holonomy in string theory and M
  theory},''
\href{http://arxiv.org/abs/hep-th/0201114}{{\tt arXiv:hep-th/0201114
  [hep-th]}}.

\bibitem{Dasgupta:1999ss}
K.~Dasgupta, G.~Rajesh, and S.~Sethi, ``{M theory, orientifolds and G -
  flux},'' \href{http://dx.doi.org/10.1088/1126-6708/1999/08/023}{{\em JHEP}
  {\bf 9908} (1999)  023},
\href{http://arxiv.org/abs/hep-th/9908088}{{\tt arXiv:hep-th/9908088
  [hep-th]}}.

\bibitem{Collinucci:2010gz}
A.~Collinucci and R.~Savelli, ``{On Flux Quantization in F-Theory},''
  \href{http://dx.doi.org/10.1007/JHEP02(2012)015}{{\em JHEP} {\bf 1202} (2012)
   015},
\href{http://arxiv.org/abs/1011.6388}{{\tt arXiv:1011.6388 [hep-th]}}.

\bibitem{Fulton}
W.~Fulton, {\em {Introduction to toric varieties}}.
\newblock Princeton University Press, Princeton, 1993.

\bibitem{Collinucci:2009uh}
A.~Collinucci, ``{New F-theory lifts. II. Permutation orientifolds and enhanced
  singularities},'' \href{http://dx.doi.org/10.1007/JHEP04(2010)076}{{\em JHEP}
  {\bf 1004} (2010)  076},
\href{http://arxiv.org/abs/0906.0003}{{\tt arXiv:0906.0003 [hep-th]}}.

\bibitem{Krause:2012yh}
S.~Krause, C.~Mayrhofer, and T.~Weigand, ``{Gauge Fluxes in F-theory and Type
  IIB Orientifolds},'' \href{http://dx.doi.org/10.1007/JHEP08(2012)119}{{\em
  JHEP} {\bf 1208} (2012)  119},
\href{http://arxiv.org/abs/1202.3138}{{\tt arXiv:1202.3138 [hep-th]}}.

\bibitem{Krause:2011xj}
S.~Krause, C.~Mayrhofer, and T.~Weigand, ``{$G_4$ flux, chiral matter and
  singularity resolution in F-theory compactifications},''
  \href{http://dx.doi.org/10.1016/j.nuclphysb.2011.12.013}{{\em Nucl.Phys.}
  {\bf B858} (2012)  1--47},
\href{http://arxiv.org/abs/1109.3454}{{\tt arXiv:1109.3454 [hep-th]}}.

\bibitem{Hayashi:2008ba}
H.~Hayashi, R.~Tatar, Y.~Toda, T.~Watari, and M.~Yamazaki, ``{New Aspects of
  Heterotic--F Theory Duality},''
  \href{http://dx.doi.org/10.1016/j.nuclphysb.2008.07.031}{{\em Nucl.Phys.}
  {\bf B806} (2009)  224--299},
\href{http://arxiv.org/abs/0805.1057}{{\tt arXiv:0805.1057 [hep-th]}}.

\bibitem{Grimm:2011fx}
T.~W. Grimm and H.~Hayashi, ``{F-theory fluxes, Chirality and Chern-Simons
  theories},'' \href{http://dx.doi.org/10.1007/JHEP03(2012)027}{{\em JHEP} {\bf
  1203} (2012)  027},
\href{http://arxiv.org/abs/1111.1232}{{\tt arXiv:1111.1232 [hep-th]}}.

\bibitem{Collinucci:2008pf}
A.~Collinucci, F.~Denef, and M.~Esole, ``{D-brane Deconstructions in IIB
  Orientifolds},'' \href{http://dx.doi.org/10.1088/1126-6708/2009/02/005}{{\em
  JHEP} {\bf 02} (2009)  005},
\href{http://arxiv.org/abs/0805.1573}{{\tt arXiv:0805.1573 [hep-th]}}.

\end{thebibliography}\endgroup
\bibliographystyle{utphys}

\end{document}